\documentclass{ws-procs9x6}            
\usepackage[utf8]{inputenc}
\begin{document}
\title{Unveiling the structure of pseudoscalar mesons}

\author{Kh\'epani Raya$^*$ and Lei Chang}

\address{School  of  Physics,  Nankai  University,  Tianjin  300071,  China\\
$^*$E-mail: khepani@nankai.edu.cn}

\author{Minghui Ding and Daniele Binosi}

\address{European Centre for Theoretical Studies in Nuclear Physics and Related Areas\\
	(ECT$^*$) and Fondazione Bruno Kessler\\
Villa Tambosi, Strada delle Tabarelle 286, I-38123 Villazzano (TN) Italy}

\author{Craig D. Roberts}

\address{School of Physics, Nanjing University, Nanjing, Jiangsu 210093, China \\
	Institute for Nonperturbative Physics, Nanjing University, Nanjing, Jiangsu 210093, China}

\begin{abstract}
A valuable approach to the analysis of hadron physics observables is provided by QCD's equations-of-motion; namely, the Dyson-Schwinger equations. Drawing from a diverse collection of predictions, we revisit: $\gamma \gamma^* \to$ neutral pseudoscalar transition form factors, their corresponding valence-quark distribution amplitudes and a recent result on the pion distribution functions. 
\end{abstract}

\keywords{Dyson-Schwinger equations, distribution amplitudes, distribution functions, transition form factors.}

\bodymatter

\section{Introduction}\label{aba:sec1}
In an exciting era of ongoing and forthcoming experiments, the Dyson-Schwinger equation (DSE) approach to QCD, which describes hadrons in terms of quarks and gluons, is placed as a promising tool for hadron physics~\cite{Horn:2016rip, Eichmann:2016yit, Burkert:2019bhp}. The present manuscript sums up a collection of DSE results on the pseudoscalar mesons: valence-quark distribution amplitudes (PDAs)~\cite{Chang:2013pq,Ding:2015rkn,Ding:2018xwy}, $\gamma^* \gamma^* \to\;M$ transition form factors (TFFs)~\cite{Raya:2015gva,Raya:2016yuj,Ding:2018xwy,Raya:2019dnh} and the pion distribution functions (PDF)~\cite{Ding:2019lwe}. 
\section{Valence-quark distribution amplitudes}
\begin{centering}
	\begin{figure}[t]
		\includegraphics[width=4.5in]{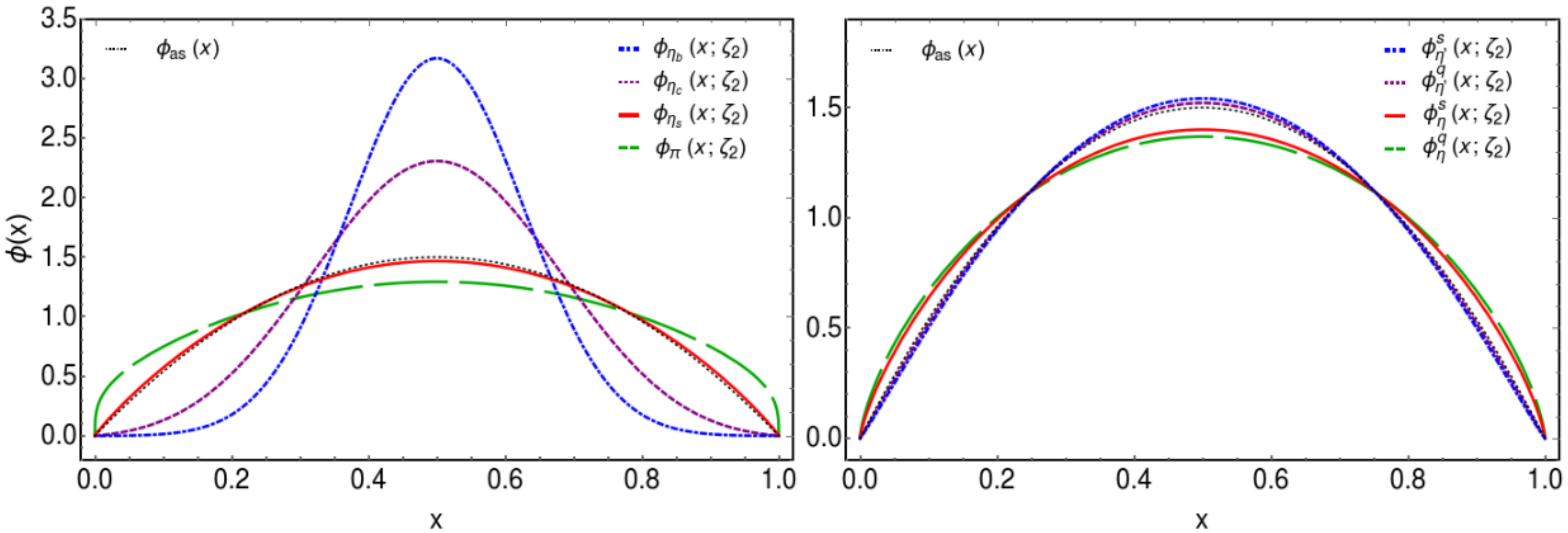}
		\caption{PDAs at $\zeta_2$ and the asymptotic profile~\cite{Lepage:1980fj}, $\phi_{\textrm{as}}(x):=6x(1-x)$. The $\eta_s$ corresponds to a pseudoscalar with $s$-massive valence-quarks. Due to the dominance of dynamical chiral symmetry breaking (DCSB), $\phi_\pi(x;\zeta_2)$ is broader than $\phi_{\textrm{as}}(x)$, while those corresponding to $\eta_{c,b}$, more influenced by the Higgs mechanism, are narrower.  Lying close to $\phi_{\textrm{as}}(x)$, we find the $\eta_s,\;\eta$ and $\eta'$ PDAs. There is an interplay between strong and weak mass generation being dominant and the $s$-quark is the boundary~\cite{Ding:2015rkn,Ding:2018xwy}.}
		\label{aba:fig1}
	\end{figure}
\end{centering}

Given a pseudoscalar meson $M$ with mass $m_M$, the PDA, $\phi_M(x;\zeta)$, is defined as the projection onto the light-front of its Bethe-Salpeter wavefunction, $\chi_M(q;P)=S(q^+)\Gamma_M(q;P)S(q^-)$, such that:
\begin{eqnarray}
\label{eq:PDA}
\phi_M(x;\zeta)&=& \frac{Z_2}{f_M} \textrm{tr}_{\textrm{CD}} \int_q \delta_n^x(q^+) \gamma_5 \gamma \cdot n \chi_M(q;P)\;,\;\int_0^1 dx\; \phi_M(x;\zeta) = 1, \\
\delta_n^x(q^+)&=& \delta(n\cdot q^+ - x \;n\cdot P)\;,\;q^+ = q + \eta P\;,\;q^- = q-(1-\eta P)\;,\nonumber
\end{eqnarray}
where $\int_q:=\int \frac{d^4q}{(2\pi)^4}$, $\eta \in [0,1]$ defines the relative momentum between the quark/antiquark,  $f_M$ is the corresponding decay constant, $Z_2$ is the quark wavefunction renormalization constant and $n$ is a light-like four vector ($n^2=0$, $n\cdot P = -m_M$). Computation of several Mellin moments allow to reconstruct the PDA~\cite{Chang:2013pq,Ding:2015rkn,Ding:2018xwy}. A selection of PDAs is shown in Fig.~\ref{aba:fig1}.
\section{$\gamma^*\gamma^* \to M$ transition form factors}
\begin{centering}
	\begin{figure}[t]
		\includegraphics[width=4.5in]{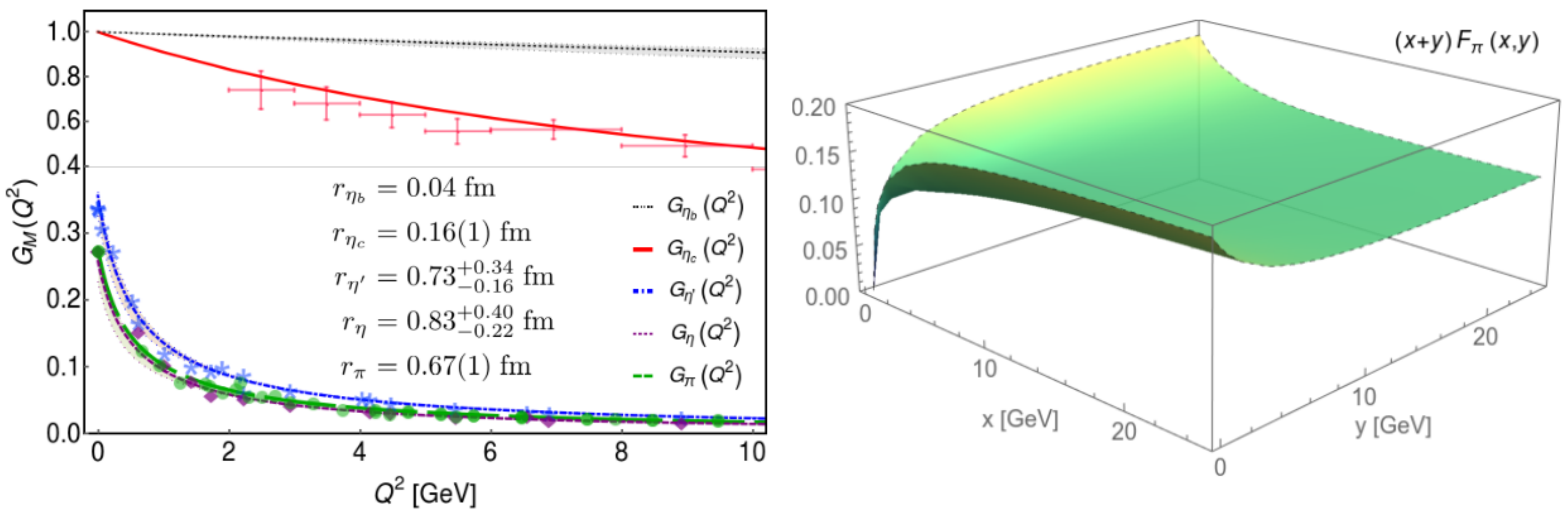}
		\caption{\textbf{Left panel}. DSE predictions of $\gamma\gamma^* \to M$ TFFs and a variety of experimental data (CELLO, CLEO, Babar, Belle, L3 and PDG) for $\eta_c$ (error bars), $\eta'$ (asterisks), $\eta$ (diamons) and $\pi^0$ (circles). See Ref.~\citenum{Raya:2019dnh} for a complete list of data sources. The $\eta_{c,b}$ results are normalized to $G_M(0)=1$ and the band surrounding the $\eta_b$ corresponds to the non-relativistic QCD result of Ref.~\citenum{Feng:2015uha}. Those form factors can be found, on a larger domain, in Refs.~\citenum{Raya:2015gva,Raya:2016yuj,Ding:2018xwy}. \textbf{Right panel}. $\gamma^* \gamma^* \to \pi^0$ TFF~\cite{Raya:2019dnh}. The ultraviolet limit constraints\cite{Lepage:1980fj} are preserved while simultaneously reproducing the empirical value at $Q^2 \to 0$.}
		\label{aba:fig2}
	\end{figure}
\end{centering}

In the impulse approximation (IA), the transition $\gamma^{*}\gamma^{*} \to M$ reads as~\cite{Raya:2019dnh}:
\begin{eqnarray}
\label{eq:defTFF}
T_{\mu\nu}(Q_1,Q_2) &=& \frac{e^2}{4\pi^2 }\epsilon_{\mu \nu \alpha \beta} Q_{1\alpha} Q_{2\beta} G_M(Q_1^2,Q_1\cdot Q_2,Q_2^2) \\ \nonumber
&=& \mathbf{e}_f^2 \; \textrm{tr}_{\textrm{CD}}\int_q i  \chi_{\mu}^f(q,q_1)\Gamma_{M}(q_1,q_2) S_f(q_2) i \Gamma_\nu^f(q_2,q)\;, 
\end{eqnarray}
where $Q_{1},\;Q_2$ are the momenta of the two photons. The kinematic arrangement is $q_1=q+Q_1$, $q_2=q-Q_2$; $\mathbf{e}_f^2$ is a charge factor associated with the meson's valence quarks. $\Gamma_M$ is the pseudoscalar Bethe-Salpeter amplitude (BSA). The amputated ($\Gamma_\nu$) and non amputated ($\chi_\mu$) quark-photon vertices can be properly expressed via the quark propagator dressing functions~\cite{Raya:2015gva}. BSAs and quark propagators are obtained (at $\zeta=\zeta_2$) in the rainbow-ladder approximation, self-consistent with the IA~\footnote{The non-Abelian anomaly is introduced in the Bethe-Salpeter kernel of the $\eta-\eta'$ BSE~\cite{Ding:2018xwy} and a mild amendment of Eq.~(\ref{eq:defTFF}) is required.}. Via perturbation theory integral representations of those objects, the complete results basically require a series of perturbation-theory-like integrals. A similar approach was followed for the pion electromagnetic form factor~\cite{Chang:2013nia}. The computed TFFs are displayed in Figs.~\ref{aba:fig2}.


\section{Pion distribution functions}
\begin{center}
	\begin{figure}[t]
		\includegraphics[width=4.5in]{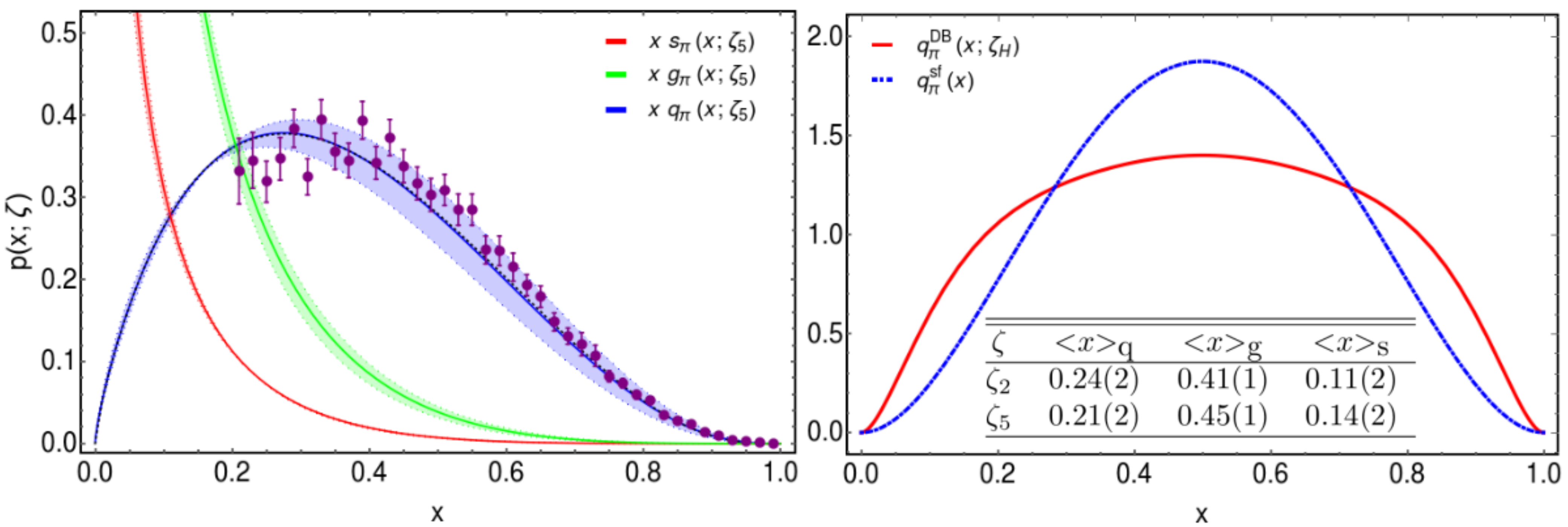}
		\caption{ \textbf{Left panel:} valence quark ($q_\pi$), gluon ($g_\pi$) and sea ($s_\pi$) pion PDFs at the experimental scale~\cite{Conway:1989fs,Aicher:2010cb}, $\zeta=5.2$ GeV $:=\zeta_5$ . \textbf{Right panel:} valence-quark PDF at $\zeta_H$. Here, $q_\pi^{\textrm{DB}}(x;\zeta_2)$  is the DSE prediction from Ref.~\citenum{Ding:2019lwe} and $q_\pi^{\textrm{sf}}(x)=30x^2(1-x)^2$ is a scale-free parton-like model~\cite{Chang:2014lva}, the marked broadening of $q_\pi^{\textrm{DB}}(x;\zeta_2)$ relative to $q_\pi^{\textrm{sf}}(x)$ is a consequence of DCSB.}
		\label{aba:fig3}

	\end{figure}
\end{center}

Detailed considerations have yielded the following symmetry-preserving expression~\cite{Ding:2019lwe,Chang:2014lva} for the pion valence-quark PDF, $q_V(x;\zeta)$:
\begin{equation}
\label{eq:PDF}
q_V(x;\zeta)= \textrm{tr}_{\textrm{CD}} \int_q \delta_n^x(q^+) n \cdot \partial_{q^+} [\Gamma_\pi(q^+;-P)S(q^+)]\Gamma_\pi(q^-;P)S(q^-)\;.
\end{equation}
The renormalization scale  $(\zeta=\zeta_H \sim 0.3 \;\textrm{GeV})$  is defined such that the fully dressed (valence-quarks) quasiparticles are the correct degrees of freedom \cite{Rodriguez-Quintero:2019fyc}, $\int_0^1dx\;q_V(x;\zeta_H)=1$. We reconstruct $q_V(x;\zeta_H)$ from its Mellin moments and evolve through DGLAP equations to incorporate gluon and sea content in the manner that QCD prescribes \cite{Ding:2019lwe,Rodriguez-Quintero:2019fyc}. This process is detailed in Ref.~\citenum{Ding:2019lwe} and the obtained PDFs are shown in Fig.~\ref{aba:fig3}.
\section{Conclusions and scope}
We have recapitulated predictions for an array of pseudoscalar mesons properties obtained within a single unifying DSE framework.  The examples chosen are illustrative, not exhaustive.  Analyses of kaon PDFs and meson GPDs and TMDs are currently underway.
\section{Acknowledgements}
KR wants to acknowledge the local organizers for their hospitality. Work supported by: Jiangsu Province \textit{Hundred Talents Plan for Professionals}. 
\bibliographystyle{unsrt}
\bibliography{bibliography}

\end{document}